\documentclass[sigconf]{acmart}

\AtBeginDocument{%
  \providecommand\BibTeX{{%
    \normalfont B\kern-0.5em{\scshape i\kern-0.25em b}\kern-0.8em\TeX}}}

\usepackage{graphicx}
\usepackage{fancybox}
\usepackage{ulem}
\usepackage{xcolor}
\begin{document}

\title{Challenges in Docker Development: A Large-scale Study Using Stack Overflow}

\author{Mubin Ul Haque}
\email{mubinul.haque@adelaide.edu.au}
\affiliation{%
  \institution{School of Computer Science, The University of Adelaide}
  \city{Adelaide}
  \country{Australia}
}
\affiliation{%
  \institution{Cyber Security Cooperative Research Centre}
}

\author{Leonardo Horn Iwaya}
\email{leonardo.iwaya@adelaide.edu.au}
\affiliation{%
  \institution{School of Computer Science, The University of Adelaide}
  \city{Adelaide}
  \country{Australia}
}
\affiliation{%
  \institution{Cyber Security Cooperative Research Centre}
}

\author{M. Ali Babar}
\email{ali.babar@adelaide.edu.au}
\affiliation{%
  \institution{School of Computer Science, The University of Adelaide}
  \city{Adelaide}
  \country{Australia}
}
\affiliation{%
  \institution{Cyber Security Cooperative Research Centre}
}

\renewcommand{\shortauthors}{Mubin Ul, et al.} 

\begin{abstract}
\textbf{Background:} Docker technology has been increasingly used among software developers in a multitude of projects. This growing interest is due to the fact that Docker technology supports a convenient process for creating and building containers, promoting close cooperation between developer and operations teams, and enabling continuous software delivery. As a fast-growing technology, it is important to identify the Docker-related topics that are most popular as well as existing challenges and difficulties that developers face.
\textbf{Aims:} This paper presents a large-scale empirical study identifying practitioners' perspectives on Docker technology by mining posts from the Stack Overflow (SoF) community.
\textbf{Method:} A dataset of $113,922$ Docker-related posts was created based on a set of relevant tags and contents. The dataset was cleaned and prepared. Topic modelling was conducted using Latent Dirichlet Allocation (LDA), allowing the identification of dominant topics in the domain.
\textbf{Results:} Our results show that most developers use SoF to ask about a broad spectrum of Docker topics including framework development, application deployment, continuous integration, web-server configuration and many more. We determined that 30 topics that developers discuss can be grouped into 13 main categories. Most of the posts belong to categories of application development, configuration, and networking. On the other hand, we find that the posts on monitoring status, transferring data, and authenticating users are more popular among developers compared to the other topics. Specifically, developers face challenges in web browser issues, networking error and memory management. Besides, there is a lack of experts in this domain. 
\textbf{Conclusion:} Our research findings will guide future work on the development of new tools and techniques, helping the community to focus efforts and understand existing trade-offs on Docker topics.
\end{abstract}




\maketitle

\section{Introduction}

Each year, Stack Overflow (SoF) conducts a survey covering everything from the developers favourite technologies to their job preferences. Docker has been ranked first in  \textit{``Most Wanted Platform''}, second in \textit{``Most Loved Platform''} and third in \textit{``Platform In Use''} in the annual survey~\footnote{Survey Results: \url{https://insights.stackoverflow.com/survey/2019##work}} with the a participation of almost 90,000 developers around the world in 2019. In fact, Docker was added for the first time in the survey of 2019, looking for developers' preference in container technologies. This survey result is a clear indication of developers' massive interest in Docker technology. In addition, the Software Engineering (SE) community has experienced significant change towards new practices for continuous delivery of high-quality software products.
This new paradigm is often described as Development and Operations (DevOps) aiming to shorten the software development life-cycle through close collaboration between developer and operation teams \cite{ebert2016devops}.
Among the various practices, the use of containers combined with the microservice-style architecture have shown great promises to the DevOps automation process \cite{kang2016container}, enabling the repacking of platforms, systems and applications into reusable building blocks -- easily created by the development teams and shipped to operations for rapid deployment. The Docker technology has led the way on containerization solutions, allowing software developers to integrate and deploy processes to conveniently package their solutions and send them to the operations team \cite{anderson2015docker}.

The rapid uptake of Docker technologies by industry has also generated a thriving community of software developers. Among the most renowned communities, Stack Overflow (SoF) became a hub to thousands of developers looking for information, asking questions and solving issues of their daily working routines. An immense amount of knowledge is created within these communities, including question and answers on Docker technology, on configuration, troubleshooting, and various experiences. In the past, various works have turned to this vast repository of knowledge, using data mining techniques to understand the software developers' perspectives on various matters, such as DevOps \cite{zahedi2020devops}, Machine Learning (ML) \cite{whatML, whyML}, Big Data \cite{bigdata}, and software security \cite{yang2016security}.

However, the mass adoption of container technologies, such as Docker, creates new challenges for software developers as it requires knowledge on various domains like networking, operating system, cloud computing and software engineering \cite{pahl2017cloud}. In order to help Docker developers, it is important to understand the topics that are most popular or that present difficulties in getting appropriate solutions. In turn, this will help to address their challenges. Without such understanding it is harder for practitioners to decide where to focus their efforts. In fact, this not only helps practitioners but also researchers and educators, providing significant contribution to the broader community.

In this paper, we present an empirical study exploring the practitioners' perspectives on Docker technology by mining posts from the SoF community.
SoF provides a rich source of data in which views and perspectives are discussed regardless of any specific organizational context.
This allows us to investigate trends and provide an overview of the current practices in regards to Docker technology. The number of Docker-related questions and users in SoF has expanded over the last seven years (specifically after 2015)
~\footnote{The distribution of posts per year has been provided in reproduction package \cite{github}}. 
This demonstrates that not only the number of developers is increasing but also software developers encounter different challenges in Docker related tasks.
To the best of our knowledge, there has not been any prior work focusing on Docker by mining question and answer (Q\&A) websites, such as the SoF. 
This study fills this gap in the literature with a large-scale investigation of posted questions, identifying dominant topics and challenges reported by practitioners.
This is done through an empirical study using Natural Language Processing (NLP) techniques for topic modelling, as well as an in-depth analysis and discussion of our findings.
This study departs from the following set of research questions:
\vspace{- 5 pt}
\begin{enumerate}
    \item [RQ1] \textbf{Docker Topics} -- What Docker topics do developers ask question about? \newline  -- The aim of this research question is to showcase the topics and categories covered by Docker-related questions. This research question can provide a deeper insight into Docker-related posts.
    \item [RQ2] \textbf{Topic Characteristics} -- What are the characteristics in terms of popularity and difficulty of the identified topics? \newline -- After identification of topics and categories, investigation has been performed to identify the most difficult and popular topics and also if they have any correlation. This research question could help to understand trends and challenges of Docker-related topics.
    \item [RQ3] \textbf{Docker Expertise} -- Do we have enough experts to address Docker related questions? \newline -- The target is to investigate whether a lack of active experts is one of the factors behind the difficulty in getting answers to Docker-related questions.
\end{enumerate}
Among the main findings, this study identifies 30 Docker-related topics that are being discussed by software developers.
These topics were further organised into 13 categories.
More than half of the posts (56.9\%) fall into four categories of (a) application development, (b) configuration, (c) networking, and (d) basic concepts. 
It is more challenging for developers to solve their issues for the topics on web browsers, networking errors and memory management.
Also, there is a substantial lack of Docker experts in the SoF community when compared to other areas such as web development. 
%
%

The rest of the paper is organized as follows. The data collection and preparation processes are described in Section \ref{sec: method}. Experimental results and discussions are presented and discussed in Section \ref{sec: results} and Section \ref{sec: implication}. Section \ref{sec: threats} discusses threats to validity. Related works are presented in Section \ref{sec: related}. Finally, the paper is concluded with future direction in Section \ref{sec: concl}.

\section{Methodology} \label{sec: method}
In the first step of analysis, we gathered the SoF dataset which is publicly available in SOTorrent \cite{sotorrent}. 
The dataset includes a large set $S$ of question and answer posts with a set of data for each post. 
Among others, the data for a post includes its identifier, its type (question or answer), title, body, tags, creation date, view count, score, favorite count and the identifier of the accepted answer for the post if the post is a question. 
An answer to a question is accepted if the contributor who posted the question marks it as accepted. Besides, a question can have at least one and at most five tags. 
Google's BigQuery \cite{bigquery} platform has been used to extract $46,947,633$ questions and answers posted over a time span of over 11 years from August 2008 to December 2019 by $4,533,602$ developer participants of SoF. 
Among these posts $18,597,996$ (39.61\%) are questions and $28,248,207$ (60.16\%) are answers of which $9,731,117$ (20.72\%) are marked as accepted answers with an acceptance rate of 34.45\%. 

\subsection{Building a Docker-related dataset}
As a first step towards answering research questions, the sub-set of SoF questions that discuss the developers Docker-related questions are identified. 
%
%

\subsubsection{\textbf{Tag-based filtering}}
In this step of analysis, a set of Docker-related tags $T$ was developed to determine and extract Docker questions from SoF. 
First, we start with a tag set $T_0$ that includes initial Docker tags. In our study, $T_0$ consists of tag $'docker'$. 
Second, we extracted questions $P$ from our dataset $S$ whose tags match a tag in $T_0$. 
Third, a set of candidate tags $T_1$ was constructed  by extracting tags of questions in $P$. 
Finally, we refined $T$ by keeping tags that are significantly relevant to Docker and excludeding others. 
We used two heuristics $\alpha$ and $\beta$ from previous work \cite{concurrency}, \cite{bigdata} and \cite{yang2016security} to measure the significance and relevance of each tag $t$ in the Docker tag set $T$.
\begin{displaymath}
     (Significance)\hspace{1.8 mm} \alpha = \frac{Number\hspace{0.8 mm}of\hspace{0.8 mm}questions\hspace{0.8 mm}with\hspace{0.8 mm}tag\hspace{1.2 mm}t \hspace{0.8 mm}in \hspace{0.8 mm}P}{Number \hspace{0.8 mm}of \hspace{0.8 mm}questions \hspace{0.8 mm}with \hspace{0.8 mm}tag \hspace{1.2 mm}t \hspace{0.8 mm}in \hspace{0.8 mm}S}
\end{displaymath}
\begin{displaymath}
     (Relevance)\hspace{1.9 mm} \beta = \frac{Number\hspace{0.8 mm}of\hspace{0.8 mm}questions\hspace{0.8 mm}with\hspace{0.8 mm}tag\hspace{1.2 mm}t \hspace{0.8 mm}in \hspace{0.8 mm}P}{Number \hspace{0.8 mm}of \hspace{0.8 mm}questions \hspace{0.8 mm}in \hspace{0.6 mm}P}
\end{displaymath}
We observed a tag $t$ to be relevant to Docker if its $\alpha$ and $\beta$ values are higher than or equal to certain thresholds. 
Our experiments used an extensive range of thresholds for $\alpha$ = \{0.1, 0.2, 0.3\} and $\beta$ = \{0.005, 0.01, 0.015\}. 
Here, we manually inspected each tag for each configuration of $\alpha$ and $\beta$ to justify its relevance to the Docker and discarded other tags which are not useful to identify Docker tags. 
For instance, consider the tag \textit{kubernetes} which occurred frequently with the Docker tags, however, it was not added since it is not exclusively used with Docker. 
The complete list of our candidate tags with different configurations can be found in our reproduction package at  \cite{github}. 
%
From our experiments, we demonstrated that for $\alpha$ = 0.1 and $\beta$ = 0.01, it provides a significantly relevant set of Docker tags. 
With these threshold values, the set $T$ of our Docker tags becomes \textit{docker, docker-compose, dockerfile, docker-swarm, docker-machine, docker-registry, boot2docker}. 
%
%

We extract all posts with at least one of such tags to develop the Docker $Post_{tag} $. 
Yang \textit{et al.} \cite{yang2016security} and Mehdi and Raffi \cite{concurrency} used similar strategy to generate their set of security and concurrency tags. 
In addition, our threshold values are also consistent with thresholds presented in prior works \cite{yang2016security} and \cite{concurrency}.

\subsubsection{\textbf{Content-based filtering}}
There might be cases in which a Docker post does not have any of these seven tags. 
One such example is the post with question id 26787241 on SoF~\footnote{\url{https://stackoverflow.com/questions/26787241/exposing-redis-db-docker-container-to-a-nodejs-docker-container}}, discussing Docker container linking issues which does not have any tags related to Docker. 
Since there is a lack of standard procedures to assign tags to SoF posts \cite{barua2014developers} and there are still Docker posts with no Docker tags, we motivated to examine the content of posts besides tags as well. 
For content-based filtering, we followed the approach of Le \textit{et al.} \cite{le2020puminer}. 
Firstly, we developed a list of keywords~\footnote{The complete list of Docker related keywords can be found at our reproduction package in \cite{github}} related to Docker inspired by the approach followed by Pletea \textit{et al.} \cite{pletea2014security}.
Then, we performed subword matching for the keywords whose length is greater than three-character to reduce false-positive as described in \cite{pletea2014security}.
Afterward, we determined the ratio of Docker words ($kw_{ratio}$) and count of Docker keywords that appear in total ($kw_{count}$). 
We used these two parameters ($kw_{ratio}$) and ($kw_{count}$) to increase the coverage of Docker discussion so that it can reduce outlier in the discussion. 
For example, post with id 56408913 on SoF has the word 'docker', however, the discussion focuses mainly the installation of \textit{yourls}.
We have obtained thresholds of $x = 0.051$ and $y = 4$ for  $kw_{ratio}$ and $kw_{count}$, respectively, from unclosed Docker posts $Post_{tag}$. 
Since a closed post may pose a possibility of being duplicated and irrelevant to discussion~\footnote{\url{https://stackoverflow.com/help/closed-questions}}, unclosed posts have been considered for ensuring Docker relevancy. 
Content-based filtering approach selects SoF posts  whose $kw_{ratio}$ $\geq$ $x$ and $kw_{count}$ $\geq$ $y$ to develop $Post_{content}$ of Docker posts which discuss about '$docker$' but does not have Docker in tag-set.

\subsubsection{\textbf{Extraction of Docker posts}}
We combined both $Post_{tag}$ and $Post_{content}$ to develop our Docker-related dataset $D$. 
The $Post_{tag}$ and $Post_{content}$ provide 69,768 and 11,522 posts, respectively.
Therefore, the total number of question-posts become 81,290. 
Afterwards, answer-posts were considered from 81,290 question posts. 
It provides 93,183 answer posts where 32,632 answer posts are considered as accepted answer posts. 
To reduce the noise, only accepted answers are taken into consideration by following earlier approaches \cite{concurrency}, \cite{bigdata} and \cite{rosen2016mobile}. 
Finally, the data set $D$ contains 113,922 SoF questions and answers of which 81,290 (71.4\%) are questions and 32,632 (28.6\%) are accepted answers.

\subsection{Data Preparation}
After building Docker-related dataset, we filtered out the irrelevant information before applying the topic modeling techniques as raw posts contain too much noise for learning a model \cite{barua2014developers}. 
In this preparation step, title, question and answers of a post are considered.

\subsubsection{\textbf{Preprocessing of Docker Posts}}

In this step, we removed code snippets marked with <code></code>, HTML tags, such as paragraph <p></p>, urls <a></a>, stop words such as 'a', 'how', 'can' by using NLTK stopwords corpus \cite{nltkstopwords} for preprocessing the set of Docker posts $D$ to minimize the noise by following previous approach \cite{yang2016security}, \cite{barua2014developers} and \cite{bigdata}. 
Afterwards, bigram models were developed by using Gensim \cite{gensim} since bigram models enhance the quality of text processing as reported by Tan \textit{et al.} \cite{tan2002use}.
In addition, we leveraged Porter stemmer \cite{porter} to reduce words to their stemmed root representations, for instance, “programming”, and “programmer” both get stemmed to word “program”).

\subsubsection{\textbf{Identifying Docker Topics and Categories}} \label{sec:topics}
Latent Dirichlet Allocation (LDA) modeling technique \cite{LDA} was adopted for this paper.
LDA has been widely used in Software Engineering studies \cite{barua2014developers}, \cite{concurrency}. 
LDA groups posts of our dataset into a set of topics based on word co-occurrences and frequencies. 
LDA assigns to each post a series of probabilities that indicate the chances of a post being related to a topic. 
The topic with the maximum probability for a particular post (i.e., the post that contains more keywords of a particular topic) is considered to be the dominant topic for a post. 

In our study, we use the MALLET~\footnote{MAchine Learning for LanguagE Toolkit (MALLET): \url{http://mallet.cs.umass.edu/topics.php}} implementation of LDA, which was adopted in previous approaches  \cite{concurrency}, \cite{bigdata}. 

\begin{figure*}[ht]
  \centering
  \includegraphics[width=1.0\textwidth]{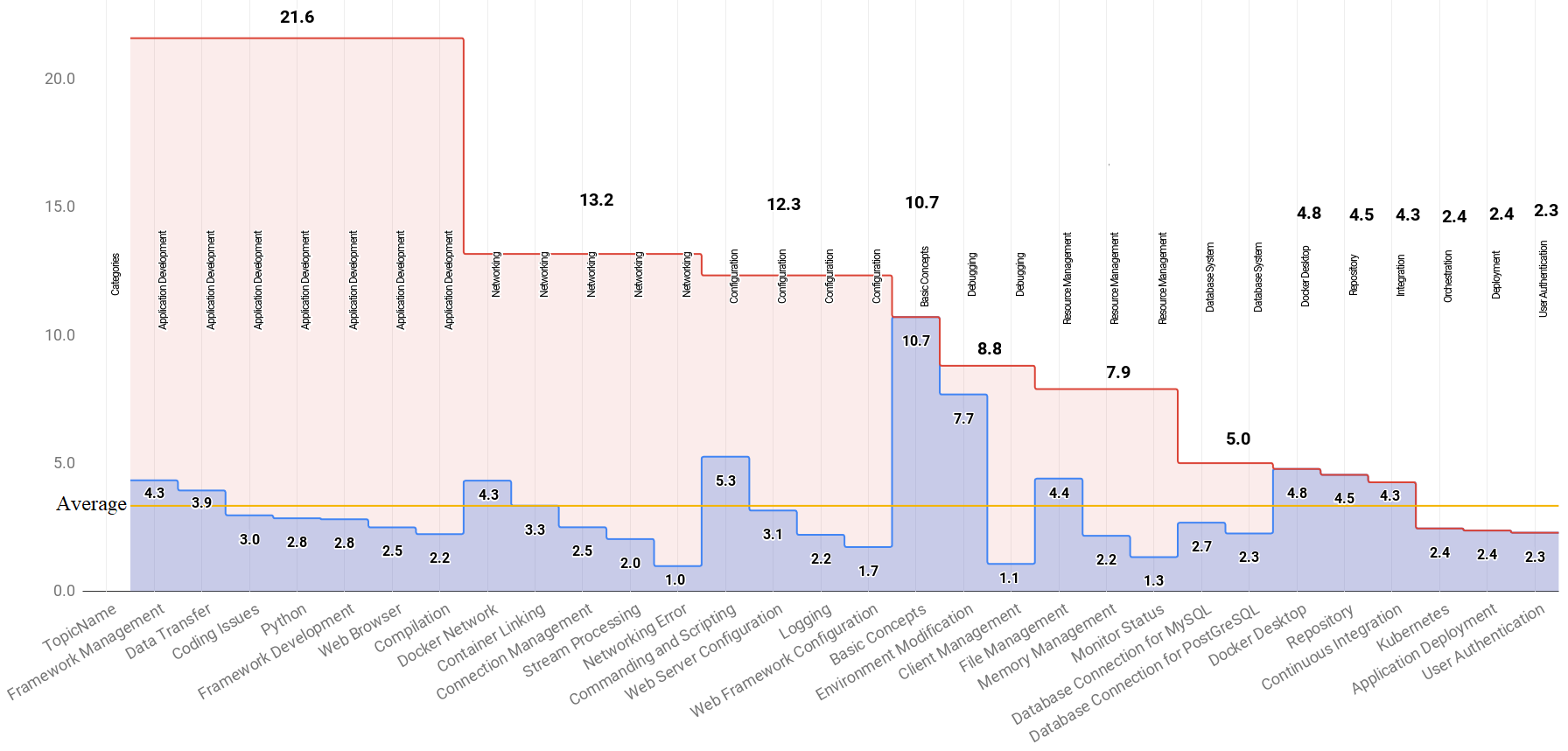}
  \caption{Distribution of Docker-related topics  and their higher level categorization (in percentage).}
  \label{fig:topic}
\end{figure*}

The main concern of implementing LDA is to determine the optimal number of topic $K$, if $K$ is too large, LDA might produce excessively fine-grained topics which are hard to analyze. 
On the other hand, LDA might yield overlapping topics when $K$ is too small. 
In this respect, we performed a broad range of experiments by varying $K$ from 5 to 50, in steps of 5, iteration value of 100, 500 and 1000 and note down the coherence score at each run. 
The coherence score measures the understandability of the LDA topics using various measures and has been shown to be highly correlated human comprehensibility \cite{roder2015exploring}. 
Hence, the first two authors run the LDA with different $K$ values and then stored each run's corresponding coherence score. 
We considered that $K$ values within the 30 to 40 range have very similar coherence scores (i.e. the variation is very small, and coherence score is  ~0.6 which is regarded as a good coherence score \cite{stevens2012exploring}). 
The first two authors tested randomly selected sample of 30 posts from each topic for $K$ values of 30 to 40 to assure that we chose the best $K$ value. 
Based on this experiment, we found that a $K$ value of 30  with 1000 iteration value provides sufficiently granular set of topics for our Docker-related dataset $D$. 
Besides, we set the hyper-parameter of LDA $\alpha$ = 50/$K$ and $\beta$ = 0.01 which are inline with the values used by earlier works \cite{bigdata}, \cite{concurrency}, \cite{yang2016security}. 
In general, MALLET only generates a list of 10-word sets as topics but it cannot infer the actual meaningful topic names.
To label topics, we used open card sort method \cite{fincher2005making} by following previous work \cite{yang2016security}, \cite{bigdata}. 
There is no predefined set of topic names in open card sort method. 
In this process, we inspected top 10 words from each topic by following guideline of Agrawal \textit{et al.} \cite{agrawal2018wrong} and read through most relevant 15 posts (based upon higher probability value for a post to belong a particular topic) for that topic to identify a topic name that best explains words and posts of that topic. 
Moreover, we also examined 15 randomly selected posts from each topic to check whether chosen topic name suits the posts. 
Similarly, we used open card sort method to cluster the similar topics into a higher level category. 
Table \ref{tab:topic} and Figure \ref{fig:topic} demonstrate 30 topic names, their topic words and categorization.
\begin{table*}[ht]
\centering
\caption{Names, categories (separated by tab) and top 10 words (stemmed topic words) for our Docker topics of SoF}
\label{tab:topic}
\resizebox{\textwidth}{!}{%
\begin{tabular}{cllc}
\toprule
\textbf{Topic\_No} & \textbf{Topic Name} & \textbf{Topic words} & \textbf{Categories}              \\
\midrule
1         & Stream Processing                  & node, cluster, kafka, connect, master, rabbitmq, messag, zookeep, worker, spark                  & Networking              \\
2         & Container Linking                  & volum, network, container\_nam, redi, environ, mongo, mongodb, link, depends\_on, restart\_alway & Networking              \\
3         & Logging                            & info, log, debug, elasticsearch, warn, info\_main, configur, fail, pid, messag                   & Configuration           \\
4         & File Management                    & volum, directori, mount, folder, data, path, copi, insid, local, share                           & Resource Management     \\
5         & Basic Concepts                     & applic, deploy, app, differ, environ, develop, possibl, base, exampl, make                       & Basic Concepts          \\
6         & Memory Management                  & memori, process, time, size, limit, set, data, cpu, number, thread                               & Resource Management     \\
7         & Networking Error                   & peer, network, order, debu, chaincod, note\_innodb, innodb, fail, channel, utc                   & Networking              \\
8         & Framework Management               & copi, project, java, applic, step, app, add, workdir, directori, expos                           & Application Development \\
9         & Docker Desktop                     & window, machin, vm, linux, ssh, default, ubuntu, connect, set, mac                               & Docker Desktop          \\
10        & Web Server Configuration           & nginx, proxi, request, locat, configur, http, root, set, app, url                                & Configuration           \\
11        & Monitor Status                     & id, latest, status, ps, minut\_ago, hour\_ago, minut, pull\_complet, pull, second\_ago           & Resource Management     \\
12        & Database Connection for PostGreSQL & databas, postgr, connect, db, postgresql, user, web\_1, data, migrat, log                        & Database System         \\
13        & Commanding and Scripting           & script, echo, bash, execut, exec, set, cmd, entrypoint, variabl, env                             & Configuration           \\
14        & Connection Management              & connect, fail, client, connect\_refus, curl, tcp, localhost, socket, request, listen             & Networking              \\
15        & Kubernetes                         & pod, kubernet, kubectl, node, deploy, cluster, spec, kubelet, apivers\_kind, minikub             & Orchestration           \\
16        & Database Connection for MySQL      & mysql, db, databas, connect, volum, link, wordpress, root, environ, user                         & Database System         \\
17        & Coding Issues                      & string, return, true, null, function, valu, code, fals, var, id                                  & Application Development \\
18        & Repository                         & registri, pull, push, tag, repositori, login, certif, authent, password, user                    & Repository              \\
19        & Environment Modification           & chang, issu, problem, updat, restart, time, remov, stop, found, solut                            & Debugging               \\
20        & Compilation                        & compil, librari, packag, make, modul, tensorflow, load, devic, model, fail                       & Application Development \\
21        & Data Transfer                      & updat, packag, env, copi, add, echo, rm, sudo, curl, mkdir                                       & Application Development \\
22        & Web Framework Configuration        & php, apach, compos, configur, directori, nginx, set, extens, exec, laravel                       & Configuration           \\
23        & Continuous Integration             & jenkin, test, stage, script, pipelin, plugin, job, step, git, gitlab                             & Integration             \\
24        & User Authentication                & user, root, sudo, fail, root\_root, permiss, group, daemon, directori, process                   & User Authentication     \\
25        & Application Deployment             & instanc, deploy, aw, task, ec, app, set, configur, azur, cluster                                 & Deployment              \\
26        & Web Browser                        & test, app, browser, web, chrome, page, url, selenium, heroku, remot                              & Application Development \\
27        & Docker Network                     & network, ip, access, ip\_address, connect, node, swarm, address, default, expos                  & Networking              \\
28        & Python                             & line, python, import, pip\_instal, return, django, app, flask, code, print                       & Application Development \\
29        & Client Management                  & fals, client, api, git\_commit, info, kernel, id, true, debug\_mode, default                     & Debugging               \\
30        & Framework Development              & npm, node, app, copi, npm\_err, nodej, yarn, depend, code, directori                             & Application Development
\\ \bottomrule
\end{tabular}%
}
\end{table*}
\section{Results and Discussions} \label{sec: results}
In this section, we present the analysis of Docker posts and topics to answer our research questions.

\subsection{RQ1. Docker Topics -- What Docker topics do developers ask question about?}
\textbf{Motivation} Docker development requires knowledge from a broad range of software engineering fields \cite{pahl2017cloud}. 
For instance, Docker developers need some knowledge in cloud computing, software engineering, operating system and distributed networking, which are often unnecessary in other conventional software development tasks. 
Therefore, difficulties faced by Docker developers are likely to vary from the traditional software development challenges.
Since developers use Questions and Answers (Q\&A) websites to interact with both problems and solutions, the purpose of this RQ is to look into the SoF's posts to identify the most prevalent Docker topics and the issues that the Docker community encounters more frequently. 
In addition, defining the commonly discussed Docker topics is the preliminary step to highlight the topics that are gaining more popularity and/or are difficult to address. 
Moreover, Docker has drawn huge attention from developers, expressing their keen interest to build application using Docker platform~\footnote{Survey Results: \url{https://insights.stackoverflow.com/survey/2019##work}}. 

%
Hence, we devise this research question to know more about their discussion topics.
\newline
\textbf{Approach} Docker topics asked by developers on SoF are determined using LDA topic inference and topic labelling as described in Section \ref{sec:topics}.
\newline
\textbf{Results} Table \ref{tab:topic} shows the topics, topic words and categories relating to Docker. 
Moreover, Figure \ref{fig:topic} demonstrates the percentage of number of topics as well as the categorization of these topics. From Figure \ref{fig:topic}, it is clear that developers ask  most questions about Basic Concepts.
Specifically, developers ask more about Framework Management, Data Transfer, Docker Network, Commanding and Scripting, Basic Concepts, Environment Modification, File Management, Docker Desktop, Repository and Continuous Integration where all of these topics have a higher percentage value than the average value for topics (3.3\%). 
All of these 30 topics are clustered together into 13 main categories. 
In following, we will explain each of the categories along with their constituent topics, sample posts from SoF for better explanation of categories and their relevancy with previous works.

\noindent\fbox{%
    \parbox{0.98\linewidth}{%
        \textbf{Finding 1: } Developers ask questions about a broad spectrum of 30 Docker topics including   Basic Concepts, Web Server Configuration, Framework Development, Application Deployment, Continuous Integration, Networking and Repository.
    }%
}

\noindent\fbox{%
    \parbox{0.98\linewidth}{%
        \textbf{Finding 2:} Docker developers' questions can be grouped into a topic hierarchy of 13 categories as Application Development, Configuration, Networking, Basic Concepts, Debugging, Resource Management, Docker Desktop, Repository, Database System, Integration, User Authentication, Orchestration and Deployment.
    }%
}

\subsubsection{\textbf{Application Development}} Application Development consists of Framework Management, Coding Issues, Compilation, Data Transfer, Web Browser, Python and Framework Development. 
This is the largest category ( approximately 21\%) among all the topics which indicates developers' concern regarding the development of applications using Docker technology. 
This concern is further strengthened by \cite{sysdig} that discussing about developers' perspective on using Dockers for various applications. 
One of the reasons behind developers' high interest in Application Development could be Docker's dominant representation in market share, which is almost 83\% as reported by \cite{melo2019using}. 
In this category, posts are related to development of Docker applications using frameworks, how these frameworks are managed and strategies for implementing specific frameworks. 
For example, developers ask about how to use Docker to develop a web application (id: \textit{48788271}, title: \textit{how to add docker support for the web application developed by using .netcore and angular5?} ), troubleshooting on building image (id: \textit{48762638}, title: \textit{"Error while building docker container with .net core"} ) and execution of application using Docker (id: \textit{37263261}, title: \textit{docker web application image needs keep running}).
\noindent\fbox{%
    \parbox{0.98\linewidth}{%
        \textbf{Finding 3:} Developers ask the most questions in Application Development and the least about User Authentication category.
    }%
}
\noindent\fbox{%
    \parbox{0.98\linewidth}{%
        \textbf{Finding 4:} Developers ask the most questions about Framework Management and the least about compilation in Application Development category.
    }%
}
\subsubsection{\textbf{Networking}} Networking category is the second largest category (approximately 13\%) in our Docker-related posts. 
Developers' attention towards Networking category is quite understandable since networking plays an important role in life-cycle of Docker containers which allows to bridge among Docker networks \cite{8250343}. 
Developers ask questions like interaction between containers  
(id: \textit{53357236}, title: \textit{"how can I communicate between containers?"}), (id: \textit{57750256}, title: \textit{"how to connect two socket.io point in different containers"} ), attach containers with Docker (id: \textit{41768157}, title: \textit{"how to link container in docker?"}) and setup network (id: \textit{58076849}, title: \textit{"how to use networking when building docker-compose?"}).
\newline
\noindent\fbox{%
    \parbox{0.98\linewidth}{%
        \textbf{Finding 5:} Developers ask more questions about Docker Network and Container Linking in Networking category.
    }%
}

\subsubsection{\textbf{Configuration}} Configuration is about configuring constituent parts of Docker applications which include Web Server Configuration, Logging, Web Framework Configuration, Commanding and Scripting topic. 
Commanding and Scripting topics form the largest share (nearly 43\%) in this category, followed by Web Server Configuration (25\%), Logging (18\%) and Web Framework Configuration (14\%). 
Some examples of posts in this category are web framework configuration  
(id: \textit{33167888}, title: \textit{"how to configure docker to use apache and php in separate containers"} ), Logging (id: \textit{54552566}, title: \textit{"how to collect logs on AWS from dockerized spring boot?"}), and Web Server Configuration (id: \textit{30151436}, title: \textit{"how to run nginx docker container with custom config?"}). 
One reason behind developers' great interest in Docker can be explained with its feature of smooth configuration of the software stack without worrying about its underlying infrastructure \cite{pahl2017cloud}.
\newline
\noindent\fbox{%
    \parbox{0.98\linewidth}{%
        \textbf{Finding 6:} Configuration category is the third largest category in our Docker-related posts.
    }%
}

\subsubsection{\textbf{Basic Concepts}} Basic Concepts cover fundamental aspects of Docker technology underlying development, deployment and its ecosystem. 
It includes questions like  (id: \textit{56420969}, title: \textit{"In docker technology, what is container format, and what is it used for to do?"} ), (id: \textit{48470051}, title: \textit{"what are the real world use cases of docker containers?"}) and (id: \textit{28089344}, title: \textit{"Docker, what is it and what is the purpose"}). 
As stated earlier, the Basic Concepts topic has the highest number of posts which is inline with general understanding that developers are still looking for primary knowledge relating to Docker \cite{sultan2019container}.\newline
\noindent\fbox{%
    \parbox{0.98\linewidth}{%
        \textbf{Finding 7:} Developers ask more than average number of questions about Basic Concepts which may be interpreted as they find the basics of Docker development still challenging.
    }%
}
\subsubsection{\textbf{Database System}} This category describes various aspects of database system in Docker environment including creation, modification, accessibility and deletion of database. 
Some examples of this category include the creation of database (id: \textit{34003404}, title: \textit{"how can i create a database on mysql docker service"}, id: \textit{58936513}, title: \textit{"How can I use an environmental variable to create a dump with PostgreSQL? I found a solution but I don't know why it doesn't work the other way"}) accessibility (id: \textit{54218824}, title: \textit{"how to access mysql docker container in spring boot docker container"}), and connection (id: \textit{55824545} ,\textit{"how to connect remote docker container mysql server"}). 
Although the performance of PostGreSQL is better than MySQL in the Docker environment \cite{8441979}, our study reveals that developers ask slightly more questions on MySQL. 
One possible explanation could be the general understanding of applying MySQL in simple web-based project where main focus is primarily on data transaction \cite{mysql}.
\newline
\noindent\fbox{%
    \parbox{0.98\linewidth}{%
        \textbf{Finding 8:} Developers ask more question about MySQL than PostGreSQL in Database System.
    }%
}
\subsubsection{\textbf{Debugging}} Debugging category consists of Environment Modification and Client Management topics where Environment Modification dominates this category (nearly 88\%). 
This category contains questions regarding identification and fixing of bugs related to the execution environment. 
Some sample posts from this category discuss environment modifications (id: \textit{30738369}, title: \textit{"Docker Neo4j Container keeps stopping - how to debug?"}), and management of client issues (id: \textit{29727171}, title: \textit{"docker-compose up leads to client and server don't have same version (client : 1.14, server: 1.12) error but client and server have the same version"}).

\subsubsection{\textbf{Resource Management}} Resource Management category consists of File Management, Memory Management and Monitor Status topics. 
This category mainly covers management, distribution and format of different resources like files and memory. 
This category includes posts like file management (id: \textit{49673250}, title: \textit{"Access files inside docker container over php"}), memory management (id: \textit{44533319}, title: \textit{"How to assign more memory to docker container"}), monitoring status (id: \textit{28759673}, title: \textit{"Get status of an ongoing docker image pull"}).

\subsubsection{\textbf{Docker Desktop}} Docker Desktop helps developers to code and containerize their applications. 
It also enables easy coordination among core components of Docker such as the Docker Engine for various platforms, e.g., Windows and Mac operating systems. 
Docker Desktop has been released in September, 2018~\footnote{Docker Desktop, \url{https://www.docker.com/blog/get-to-know-docker-desktop/}}. 
Developers ask about various questions on different platforms such as  Windows (id: \textit{48376928}, title: \textit{"On Windows Setup, how can I get docker image from local machine"}), Mac operating system issues (id: \textit{40007657}, title: \textit{"OS X - how to give permissions to jenkins user to run docker command"}), and Ubuntu (id: \textit{33990483}, title: \textit{"Docker, how to add a folder in OS Ubuntu in Container"}).

\subsubsection{\textbf{Repository}} Repository refers to the storage of Docker images where developers can upload or download images for using in their development areas. 
Docker Hub~\footnote{Docker Hub: \url{https://hub.docker.com}} is the largest storage for image repositories. 
It contains more than 100,000 container images provided by various software vendors, open source community and projects. 
Repositories contain the entire ecosystem to run a target software including its configuration details, operating system related utilities, packages and libraries \cite{xu2018mining}.
Developers ask questions about uploading image to a repository (id: \textit{31453639}, title: \textit{"Docker api: push an image to docker hub private repository"}), and providing a tag to an image (id: \textit{37134929}, title: \textit{"how to tag image in docker registry"}).

\subsubsection{\textbf{Integration}} Integration means aggregation of several sub-components of a software system into one system. 
Integration using Docker has become crucial in continuous software engineering process.
Mansooreh \textit{et al.} presented that developers are incorporating Docker in their continuous integration process \cite{zahedi2020devops}. 
Some examples of posts in this category include, integration using specific tool (id: \textit{34306354}, title: \textit{"Integrating Docker with jenkins for continuous integration"}), and simulating integration tests (id: \textit{57472797}, title: \textit{"Testcontainers - Run container with different database in integration test"})

\subsubsection{\textbf{Orchestration}} Orchestration simplifies the management of container clusters and their life-cycle. 
Docker supports various orchestration tools such as Mesos\cite{hindman2011mesos}, Docker Swarm~\footnote{Docker Swarm: \url{https://docs.docker.com/engine/swarm/}}, and Kubernetes~\footnote{Kubernetes: \url{https://kubernetes.io}}. 
Kubernetes is used extensively in comparison with other tools as indicated by \cite{pan2019performance} due to its compatibility with any operating system, auto-scaling feature for large deployments and better resource management facilities. 
Developers in this category have queries about using Kubernetes in Docker (id: \textit{39995335}, title: \textit{"access docker container in kubernetes"}), and initialization of Kubernetes in Docker (id: \textit{32955684}, title: \textit{"How to pass a parameter to docker with kubernetes"}).
\newline
\noindent\fbox{%
    \parbox{0.98\linewidth}{%
        \textbf{Finding 9:} Developers ask more question about Kubernetes than Docker Swarm in Orchestration tool.
    }%
}

\subsubsection{\textbf{Deployment}} Docker has been designed for easy creation, deployment, and execution of desired applications with the help of containers \cite{garg2019automated}.
As containers can be configured with all the required packages (and dependencies) needed to run an application, deployment can be performed as a package. 
Some examples of posts in this category are the deployment in cloud platforms (id: \textit{39083768}, title: \textit{"AWS Docker deployment"}), and deployment using native commands (id: \textit{48421169}, title: \textit{"how to deploy specific docker container just by docker run?"})

\subsubsection{\textbf{User Authentication}} User authentication refers to a set of posts that discuss about allowing a user or entity to verify its identity or credentials in order to access Docker resources and functionalities. 
Generally, it requires authorization to create or edit container resources, network configuration, deploy or develop image and integrate application instances. 
Developers ask questions in regards to approval issues while using a web-server (id: \textit{47593357}, title: \textit{"Facing authentication and Permissions issue when building nginx container"}), and the process of identification of users in Docker (id: \textit{56511320}, title: \textit{"Docker new user on host, how to let docker know?"}).
%


\subsection{RQ2. Topic Characteristics –- What are the characteristics in terms of popularity and difficulty of the identified topics?}
\textbf{Motivation} This RQ focuses on the popularity and difficulty of question for each topic in a more fine-grained manner. 
Identifying popular and difficult topics allows us to organise the existing knowledge base and draw concrete insights.
That is, we are able to identify topics that are trending in the Docker community as well as the adoption of specific technologies. 
Also, difficult topics suggest that more attention might be needed to improve frameworks and tools of which developers are facing most challenges during their use or configuration.
We also examine if there is any significant statistical correlation between popular and difficult topics in order to characterise Docker-related posts. 
\newline
\textbf{Approach} We use the average number of \textit{views}, \textit{favorite count} and \textit{score} of a question to measure the popularity of topics. 
Similarly, the percentage of questions of a topic that has \textit{no accepted answers} and average median \textit{time needed for questions of a topic to receive accepted answers} have been used to determine the difficulty of a topic. 
These 5 metrics are all well-known in the literature and were also used in \cite{yang2016security}, \cite{bigdata}, \cite{concurrency}, \cite{rosen2016mobile}.
The average number of views (\textit{View}) of the question indicates the interest of the community. 
If a question received a large number of views (from both registered and unregistered users as Q\&A sites like SoF have more unregistered users who visit the site frequently \cite{mamykina2011design}), then the question can be inferred as a popular question among developers. 
Besides, the average number of favorites (\textit{Favorite}) of a question measures the issues and solutions that developers found helpful and that are likely to be recurring questions during Docker development process. 
The average score (\textit{Score}) of a question can be interpreted as usefulness to its users, where a user can up-vote a post and these votes are summed up to define the score.
Moreover, the author of a question can mark an answer as an accepted answer if the answer has the merit to satisfy and solve the original question. 
In that respect, we measure percentage of questions that have no accepted answers for each topic. 
The topics which received fewer accepted answers can be interpreted as difficult topics for the developers. 
We also calculate the median time (in minutes) to receive an accepted answer for a question.
The longer the time to receive an accepted answer can be explained as a more difficult question \cite{whyML, whatML, concurrency, bigdata}. 
In addition, we use Kendall correlation \cite{abdi2007kendall} to identify and verify correlation between popular and difficult topics. 
We choose Kendall correlation since it is less sensitivity to outliers, thus producing more robust results \cite{puth2015effective}.
\newline
\textbf{Results} Table \ref{tab:popular} and \ref{tab:difficult} show popular and difficult Docker topics. Table \ref{tab:popular} has been sorted based on the average view (\textit{View}), and the topics in Table \ref{tab:difficult} are ordered based on percentage of accepted answers that they received. 
Monitor Status, Data Transfer and User Authentication are popular topics. 
For the Monitor Status topic, which has most average view and score, users tend to ask how to access Dockerfile of an image, how to save all Docker images and copy to another machine, and how to get status of a running Docker pull process. 
Monitor Status and Data Transfer topics also have a fast response time in terms of accepted answer, i.e., 79 and 105 minutes, respectively. 
The average median time to get an accepted answer is 306.8 minutes which can be interpreted as its usefulness and efficacy to the developers. 
On the other hand, Web Browser, Networking Error and Memory Management are the most difficult topics based on their percentage of non-accepted answers. 
In particular, Networking Error topic has the longest time to receive an accepted answer. 
In addition, it has the lowest score from users, which can be inferred as a challenging topic. 
In the Web Browser topic, developers generally ask about how to use a web driver when using docker (id: \textit{36773167}, title: \textit{“How do you set up selenium grid using docker on windows?”}) or why a web driver does not work in a particular platform  (id: \textit{49289050}, title: \textit{“RSelenium using Docker and Firefox, Browser Not Opening on Mac”}).

\begin{table}[ht]
  \caption{Popularity of Docker topics}
  \label{tab:popular}
  \scalebox{0.8}{
  \begin{tabular}{llll}
    \toprule
    \textbf{Topic\_Name} & \textbf{View} & \textbf{Favorite} & \textbf{Score} \\
\midrule
Monitor Status                     & 3777.35      & 3.92        & 4.73     \\
Data Transfer                      & 3566.54      & 3.81        & 3.82     \\
User Authentication                & 3084.92      & 3.63        & 3.39     \\
Commanding and Scripting           & 2943.27      & 3.55        & 3.82     \\
File Management                    & 2799.47      & 3.55        & 3.96     \\
Environment Modification           & 2725.24      & 4.20        & 3.90     \\
Client Management                  & 2681.09      & 2.85        & 3.28     \\
Connection Management              & 2319.53      & 2.92        & 2.45     \\
Docker Network                     & 2229.29      & 3.44        & 2.78     \\
Docker Desktop                     & 2131.75      & 2.84        & 2.68     \\
Database Connection for PostGreSQL & 2080.29      & 3.18        & 2.55     \\
Repository                         & 2028.69      & 2.39        & 2.38     \\
Basic Concepts                     & 1840.39      & 5.00        & 3.95     \\
Database Connection for MySQL      & 1800.25      & 2.19        & 1.68     \\
Web Framework Configuration        & 1581.98      & 1.74        & 1.62     \\
Memory Management                  & 1551.15      & 2.50        & 2.45     \\
Framework Management               & 1484.50      & 2.05        & 1.97     \\
Container Linking                  & 1438.93      & 2.06        & 1.75     \\
Framework Development              & 1397.56      & 1.84        & 1.75     \\
Kubernetes                         & 1297.48      & 1.61        & 1.39     \\
Continuous Integration             & 1279.83      & 2.08        & 1.90     \\
Python                             & 1168.78      & 1.73        & 1.64     \\
Web Server Configuration           & 1126.36      & 1.67        & 1.26     \\
Logging                            & 1077.70      & 1.87        & 1.42     \\
Compilation                        & 1045.27      & 2.04        & 1.63     \\
Web Browser                        & 897.95       & 2.18        & 1.50     \\
Stream Processing                  & 843.65       & 1.50        & 1.20     \\
Application Deployment             & 811.84       & 1.77        & 1.66     \\
Networking Error                   & 685.06       & 1.54        & 0.89     \\
Coding Issues                      & 604.06       & 1.32        & 1.05     \\
\midrule
Average                            & 1810.01      & 2.57        & 2.35  \\  
\bottomrule
\end{tabular}
}
\end{table}

\noindent\fbox{%
    \parbox{0.98\linewidth}{%
        \textbf{Finding 10:} Monitor Status in Resource Management category and Coding Issues in Application Development category are among the most and least popular Docker topics.
    }%
}

\noindent\fbox{%
    \parbox{0.98\linewidth}{%
        \textbf{Finding 11:} Web Browser in Application Development category and Commanding and Scripting in Configuration category are among the most and least difficult Docker topics.
    }%
}

As mentioned, Monitor Status and Data Transfer are among the most popular topics, even though their difficulty level is at bottom. 
Nonetheless, Web Browser and Networking Error issues rank top in difficulty but lower in popularity metrics. 
This evidence could suggest that there might be a correlation between popularity and difficulty metrics. 
To perform the correlation analysis, we measure six correlations between each of our three popularity and difficulty metrics. 
For each correlation, we identify the \textit{p-value} and co-efficient by using Kendall correlation which is shown in Table \ref{tab:correlation}. In Table \ref{tab:correlation}, the result is presented in \textit{x/y} format, where $x$ denotes co-efficient and $y$ denotes \textit{p-value}.  
We found a statistically significant correlation with 95\% confidence level between the popularity and difficulty metrics, since all correlations have \textit{p}-\textit{value} are less than $0.05$. 
In other words, difficult topics are less popular among developers, and vice-versa.
%
%

\noindent\fbox{%
    \parbox{0.98\linewidth}{%
        \textbf{Finding 12:} There is a statistically significant negative correlation between popular and difficult Docker topics.
    }%
}

\begin{table}[ht]
  \caption{Difficulty of Docker topics}
  \label{tab:difficult}
  \scalebox{0.8}{
  \begin{tabular}{lll}
    \toprule
    \textbf{Topic\_Name} & \begin{tabular}[c]{@{}l@{}}\textbf{\% w/o acc.} \\ \textbf{answer}\end{tabular} & \begin{tabular}[c]{@{}l@{}}\textbf{Mins. to}\\ \textbf{acc. answer}\end{tabular} \\
\midrule
Web Browser                        & 69.47              & 727.5                \\
Networking Error                   & 66.29              & 931                  \\
Memory Management                  & 65.24              & 308                  \\
Docker Network                     & 65.19              & 290                  \\
Logging                            & 64.92              & 641.5                \\
Compilation                        & 64.75              & 492                  \\
Web Server Configuration           & 64.01              & 334                  \\
Continuous Integration             & 63.30              & 339                  \\
Docker Desktop                     & 63.02              & 211.5                \\
Stream Processing                  & 62.96              & 502                  \\
Repository                         & 61.76              & 406                  \\
Connection Management              & 61.75              & 268.5                \\
Client Management                  & 61.55              & 608                  \\
Python                             & 61.07              & 131                  \\
Container Linking                  & 59.99              & 138.5                \\
Coding Issues                      & 59.72              & 374.5                \\
User Authentication                & 59.24              & 198                  \\
Application Deployment             & 59.20              & 532                  \\
Web Framework Configuration        & 58.99              & 162                  \\
Kubernetes                         & 58.97              & 347                  \\
Basic Concepts                     & 58.85              & 168                  \\
Framework Management               & 58.66              & 145                  \\
Database Connection for PostGreSQL & 58.41              & 165.5                \\
Framework Development              & 58.13              & 145                  \\
Environment Modification           & 57.63              & 151                  \\
Database Connection for MySQL      & 57.15              & 114.5                \\
File Management                    & 56.84              & 120                  \\
Data Transfer                      & 53.91              & 105                  \\
Monitor Status                     & 50.60              & 79                   \\
Commanding and Scripting           & 50.21              & 69                   \\
\midrule
Average                            & 60.39              & 306.8           \\
\bottomrule
\end{tabular}
}
\end{table}

\begin{table}[ht]
  \caption{Correlations of popularity and difficulty metrics}
  \label{tab:correlation}
  \scalebox{0.8}{
  \begin{tabular}{llll}
    \toprule
    \textbf{co-efficient/p-value} & \textbf{AvgView} & \textbf{AvgFav} & \textbf{AvgScore}       \\
    \midrule
    \%w/o acc. answer             & -0.4253/0.0008 & -0.2781/0.0314 & -0.3655/0.0042 \\
    Mins to acc. answer      & -0.4602/0.0004 & -0.3590/0.0054 & -0.4096/0.0015 \\
    \bottomrule
    \end{tabular}
  }
  \vspace{- 0.1 in}
\end{table}

\subsection{RQ3. Docker Expertise -- Do we have enough experts to address Docker related questions?}
\begin{figure}[ht]
  \centering
  \includegraphics[width =0.4\textwidth]{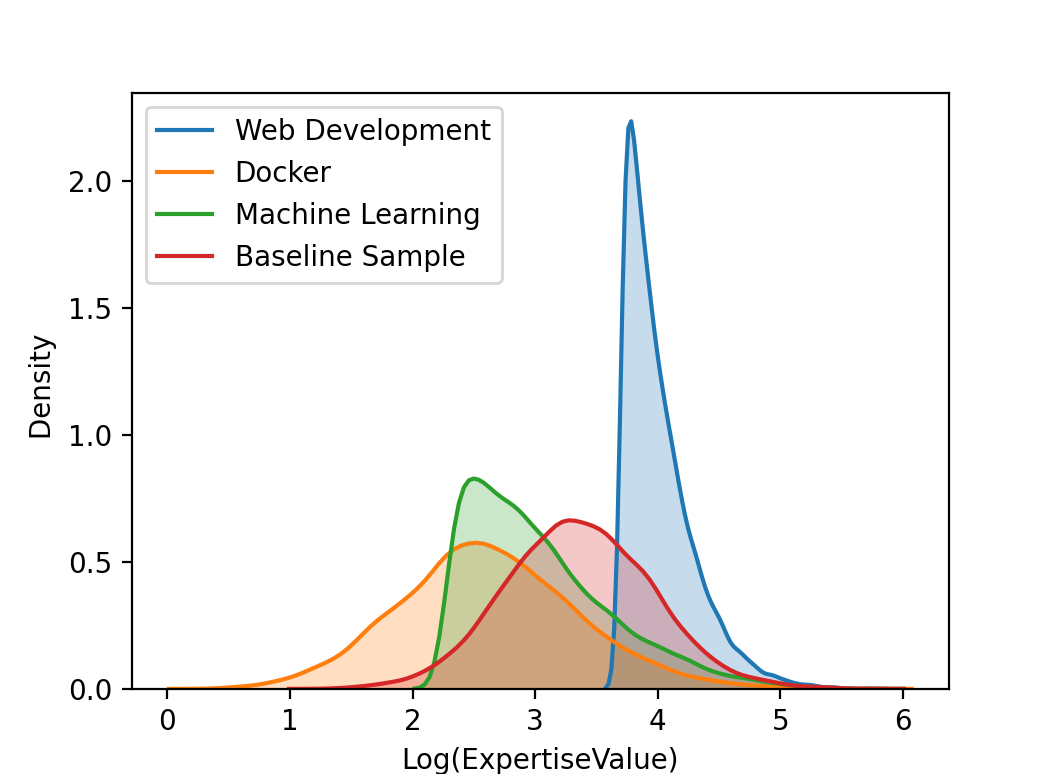}
  \caption{ Distribution of experts in Baseline Sample, Machine Learning, Web Development and Docker.}
  \label{fig:expert}
\end{figure}

\textbf{Motivation} The average percentage of no accepted answers for Docker-related posts is rather large, i.e., almost 60\% as shown in Table \ref{tab:difficult}. 
Asaduzzaman \textit{et al.} reported that the absence of adequate experts in SoF may trigger such events \cite{asaduzzaman2013answering}. 
Therefore, we investigate whether the lack of experts is one of the factors behind the difficulty in getting answers to Docker questions. By experts, we mean the users who provide significant contributions, effective and helpful answers to the SoF community \cite{cmc.2019.07818}.
\newline
\textbf{Approach} To answer this research question, we followed the approach by Yijun \textit{et al.} \cite{cmc.2019.07818}. 
In this approach, the user answers and profile performance are derived from SoF dataset and then combine together to form an \textit{ExpertiseValue} for each SoF user.
We choose this method as it considers various user attributes (for example, user reputation, quality of the answered question, profile view count of a user) and platform related information (like $tags$). 
We compare Docker domain with a Baseline Sample extracted from SoF. This sample has been generated by randomly selected over 100, 000 posts (sample size near to our Docker-related dataset). This sample size covers a broad range of $tags$ like programming (java, C\#), services (excel, gmail), databases (oracle, sql), mobile development (android, ios). In addition, Web Development is considered due to its popularity in SoF as reported in \cite{barua2014developers}. Besides, we considered the Machine Learning domain as well.
To answer RQ3, we select $tags$ derived from our tag-based filtering and same number of users from sample size along with our Docker domain (more than 12,000 users~\footnote{The complete list of users for Docker, Web Development, Machine Learning and Baseline Sample along with their $ExpertiseValue$ can be found at our reproduction package\cite{github}}). 
\newline
\textbf{Results} The comparison of experts in Docker and other domains is presented in Figure \ref{fig:expert}.
Due to skewed distribution of \textit{ExpertiseValue}, a logarithmic transformation on the \textit{ExpertiseValue} have been applied for better visualization of the distributions.
We can observe that docker experts are represented mostly at the left side of the graph which represents the presence of novice users with lower \textit{ExpertiseValue} i.e, users with less score in \textit{ExpertiseValue}. 
Besides, the right side of the graph illustrates the expert users with higher \textit{ExpertiseValue}. The users from the Baseline Sample, Machine Learning and Web Development are mostly represented here. 
In addition, a higher peak is observed for all other domain than Docker users. The higher peak indicates that the number of Docker experts is indeed smaller than in the Baseline Sample.  
Therefore, it is evident that the Docker domain indeed has a much lower number of experts than the Baseline Sample.\newline
\noindent\fbox{%
    \parbox{0.98\linewidth}{%
        \textbf{Finding 13:} There is a lack of experts in Docker domain.
    }%
}
%
\vspace{- 0.2 in}

\section{Implications} \label{sec: implication}
The findings of our study can benefit software developers, researchers and educators, both in terms of providing an overview of the Docker-related discussion as well as informing and supporting decision-making processes.
The identified topics and categories are presented in Table \ref{tab:topic}. These topics and categories usually involve a combination of domains like distributed networking, operating system and software engineering, as observed by previous researchers in \cite{pahl2017cloud}. Thus, it can be implied that Docker development also requires substantial knowledge on various domains as stated above. In addition, we identify the trends of Docker topics to better understand ongoing practices. Table \ref{tab:trends} shows the increasing, decreasing and static trends of Docker topics. This is done using Mann-Kendall test \cite{Hussain2019pyMannKendall}. We chose Man-Kendall test as it is not affected by the length of time-series \cite{Hussain2019pyMannKendall}. Note that, to determine if each topic is increasing or decreasing over time, to a statistically significant degree, we use the standard 95\% confidence level. Moreover, we used relative measure where the total number of posts for a specific topic is compared to the total number of posts for the entire Docker dataset for each month. 17 topics (nearly 57\%) have showed increasing trends. Interestingly, difficult topics such as Web Browser, Networking Error and Logging show increasing trends which represents the developers' growing interest in such topics, and researchers can focus their efforts in these topics. We also found that 4 topics have static (neither decreasing or increasing to a significant degree) trends and 9 topics (30\% of topics) have decreasing trends. Topics with decreasing trends are not outdated rather they represent a matured state.

\begin{table}[]
\centering
\caption{Analysis of Trends (ordered by Trend)}
\label{tab:trends}

\scalebox{0.8}{
\begin{tabular}{lccll}
\toprule
\textbf{Topic} & \textbf{Trend} & \textbf{Hyp.} & \textbf{p-value} & \textbf{z-score} \\
\midrule
Continuous Integration             & $\uparrow$ & T          & 0        & 10.2262 \\
Python                             & $\uparrow$ & T          & 0        & 9.4698  \\
Container Linking                  & $\uparrow$ & T          & 0        & 9.4251  \\
Kubernetes                         & $\uparrow$ & T          & 0        & 9.2630  \\
Framework Management               & $\uparrow$ & T          & 0        & 9.2139  \\
Coding Issues                      & $\uparrow$ & T          & 2.22E-16 & 8.2812  \\
Networking Error                   & $\uparrow$ & T          & 2.22E-16 & 8.2695  \\
Database Connection for MySQL      & $\uparrow$ & T          & 4.88E-15 & 7.8283  \\
Framework Development              & $\uparrow$ & T          & 1.95E-14 & 7.6537  \\
Web Server Configuration           & $\uparrow$ & T          & 8.50E-14 & 7.4621  \\
Web Browser                        & $\uparrow$ & T          & 1.63E-11 & 6.7354  \\
Logging                            & $\uparrow$ & T          & 7.23E-10 & 6.1609  \\
Database Connection for PostGreSQL & $\uparrow$ & T          & 7.23E-10 & 6.1609  \\
Web Framework Configuration        & $\uparrow$ & T          & 7.35E-10 & 6.1584  \\
Compilation                        & $\uparrow$ & T          & 7.05E-07 & 4.9601  \\
Application Deployment             & $\uparrow$ & T          & 5.59E-05 & 4.0297  \\
Stream Processing                  & $\uparrow$ & T          & 0.0028   & 2.9887  \\
Memory Management                  & --   & F          & 0.6035   & 0.5194  \\
Client Management                  & --   & F          & 0.7700   & 0.2924  \\
Repository                         & --   & F          & 0.5979   & -0.5274 \\
Connection Management              & --   & F          & 0.0899   & -1.6959 \\
File Management                    & $\downarrow$ & T          & 0.0035   & -2.9230 \\
User Authentication                & $\downarrow$ & T          & 0.0010   & -3.2776 \\
Data Transfer                      & $\downarrow$ & T          & 0.0004   & -3.5638 \\
Docker Network                     & $\downarrow$ & T          & 4.31E-05 & -4.0903 \\
Monitor Status                     & $\downarrow$ & T          & 8.82E-06 & -4.4443 \\
Commanding and Scripting           & $\downarrow$ & T          & 3.13E-08 & -5.5337 \\
Docker Desktop                     & $\downarrow$ & T          & 5.19E-09 & -5.8410 \\
Environment Modification           & $\downarrow$ & T          & 1.13E-13 & -7.4252 \\
Basic Concepts                     & $\downarrow$ & T          & 0        & -9.9892 \\ \bottomrule
\multicolumn{5}{{p{0.5\textwidth}}}{\textbf{Note:} '$\uparrow$', '$\downarrow$', '--', 'T' and 'F' denote increasing, decreasing, static, True and False respectively. \textbf{Hyp.:} Hypothesis.}
\end{tabular}%
}
\vspace{ - 0.25 in}
\end{table}

\begin{figure}[ht]
  \centering
  \includegraphics[width = 0.45\textwidth]{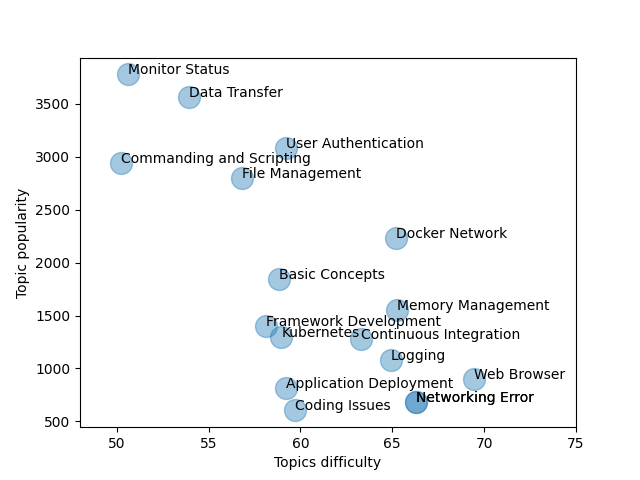}
  \caption{Trading off Docker topics (popularity vs difficulty).}
  \label{fig:correlation}
  \vspace{ - 0.15 in}
\end{figure}

In addition, Figure \ref{fig:correlation} shows the trade-off that Docker topics have in regards to their popularity and difficulty.
Based on Figure \ref{fig:correlation}, developers can choose to focus on topics that match their current skill levels. 
For instance, novice developers may benefit from learning about topics that have already been heavily discussed in the SoF community and issues that have already been solved, such as Monitor Status and Data Transfer which also show decreasing trends.
On the other hand, experienced developers might want to become more knowledgeable by tackling topics that have an increasing difficulty level, while still popular, such as Docker Network.
Managers can also use such results for assigning tasks that match their teams' skill level, e.g., avoiding giving a junior developer a very difficult and perhaps unsolved task such as Web Browser (this topic discuss about testing applications in different web browsers).
Docker developers might benefit from understanding the long-standing issues in the SoF community, so as to improve frameworks and tools that facilitate the developers' job when using and configuring Docker-related technology. 
Likewise, academic researchers may want to delve into such problematic topics in order to have a deeper understanding of topics that are challenging to developers. 
Educators that teach subjects related to Docker can also leverage from such trade-off analysis and evidence of deficiency of experts as shown in Figure \ref{fig:expert}.   
For instance, by selecting appropriate topics that should be part of a course program, as well as considering an appropriate time-frame for exercises and assignments according to the difficulty/popularity/trend of the chosen topic.
%

\section{Threats to Validity} \label{sec: threats}
We acknowledge that this study has limitations. The threats to its validity are herein discussed.
First, this study relies on questions and answers from the SoF posts as a proxy to software developers' interests, difficulties and discussions with respect to Docker.
Although SoF is provides a rich repository of knowledge for empirical investigations in software engineering, additional studies still need to be conducted in order to corroborate or generalise findings.
%

Second, identifying the Docker-related posts in the entire SoF corpus of posts can be also a limiting factor.
That is because only using tags is not enough to completely identify the set of Docker-related posts (i.e., some posts are not tagged or miss the 'docker' tag).
To minimise this threat we not only used techniques for reliable tag-based filtering but also performed content-based filtering.
Such techniques have already been used in previous work \cite{rosen2016mobile, yang2016security, concurrency, le2020puminer} enabling us to develop our Docker tags and solid experiments with a broad range of tag relevance and significance thresholds $\alpha$ and $\beta$.
%
Also, for parsing and topic modelling we used various techniques for data pre-processing and cleaning (e.g., remove stop-words, stemming and lemmatisation), as well as used established methods to derive topics using MALLET~\footnote{MAchine Learning for LanguagE Toolkit (MALLET): \url{http://mallet.cs.umass.edu/topics.php}} for LDA.

Third, although topic modelling has been proven useful for analysing large volumes of textual data, the process of manual labeling (i.e., naming the topic 10-word sets) is subject to biases of human judgement.
To minimize this threat, we followed two strategies: 
(a) two researchers performed the labelling and disagreements were discussed and resolved; 
(b) we also manually reviewed 15 posts highly related to each topic for helping during the labelling definition.
%

Fourth, determining the optimal value of K during the topic modelling process is also a potential threat.
This threat was minimised by following a common approach \cite{stevens2012exploring} of experimentation of a broad range of values for K and observation of the perplexity and coherence metrics.
Lastly, measuring popularity and difficulty of topics can also be a threat, so we also adopted known heuristics already used in previous works for calculating these metrics \cite{concurrency, bigdata}.

\section{Related Work} \label{sec: related}

As aforementioned, many papers have investigated the software developers' perspectives by mining software repositories (eg, Github) and Q\&A websites (eg, Sof). Some studies which leverage SoF data on different domains such as Machine Learning \cite{whatML, whyML}, Mobile \cite{rosen2016mobile}, Security \cite{le2020puminer, yang2016security}, Big data \cite{bigdata}, Web development \cite{bajaj2014mining} and Concurrency \cite{concurrency}. Melo \textit{et al.} used Docker to reproduce codes derived from SoF posts. Xu \textit{et al.} discussed the opportunity of mining configuration related data from image repositories like Docker Hub. 

However, existing literature does not address the Docker technology specifically.
Existing research that addresses somewhat related topics are the works on DevOps \cite{zahedi2020devops} and on Dockerfiles \cite{cito2017dockerfiles}.

The work of \cite{zahedi2020devops} focuses on the broad area of continuous software engineering, mining the SoF dataset and performing topic modelling and a qualitative analysis of existing posts.
In this work, authors identified 32 topics of discussions, among which ``Error messages in Continuous Integration/Deployment'' and ``Continuous Integration concepts'' are the most dominant.
However, this work was intentionally designed to be technology agnostic, and authors have explicitly removed tags such as 'Jenkins' and 'Docker' during the dataset creation.
Therefore, our work is more specific, targeting one of the most widely used technologies in the DevOps community, i.e., Docker containers.

The work of \cite{cito2017dockerfiles} is focused on the quality and evolution of Dockerfiles, i.e., the specific files that define an environment that aim to enable reproducible builds of the containers.
These Dockerfiles were retrieved from project repositories on GitHub.
Results show that main quality issues in Dockerfiles arise from missing/improper version pinning in 28,6\% of the cases, and irreproducible builds in 34\% of Dockerfiles from a representative sample of the projects.
Although these findings are important to software engineers, the scope of the work is rather limited to the building process. Our work is different from the above studies as we perform an empirical study on Docker-related posts on SoF. 

\section{Conclusion and Future Work} \label{sec: concl}
This paper provides a large-scale empirical study over 113,922 posts from the SoF community to understand interests and challenges faced by software developers with respect to Docker technology.
Using a topic modelling approach, a total of 30 topics were derived from the dataset of Docker-related posts and these topics were grouped in 13 main categories. These identified categories and topics are inline with use cases of Docker such as development, deployment, configuration and integration.  
The most popular topics include Monitor Status, Data Transfer and User Authentication, while the topics on Web Browser, Networking Error and Memory Management are more challenging to the community. This study also reveals that topics' difficulty and popularity are in fact negatively correlated, i.e., hard topics are less popular and vice-versa. Moreover, we found that the Docker community has a significantly smaller number of expert developers when compared to broader communities. 
%

%

\begin{acks}
 The work has been supported by the Cyber Security Research Centre Limited whose activities are partially funded by the Australian Government’s Cooperative Research Centres Programme.
This work has also been supported with super-computing resources provided by the Phoenix HPC service at the University of Adelaide. We would like to thank Triet H. M. Le, Roland Croft and anonymous reviewers for their valuable comments.   
\end{acks}

\bibliographystyle{ACM-Reference-Format}
\bibliography{main}


\begin{thebibliography}{44}


\ifx \showCODEN    \undefined \def \showCODEN     #1{\unskip}     \fi
\ifx \showDOI      \undefined \def \showDOI       #1{#1}\fi
\ifx \showISBNx    \undefined \def \showISBNx     #1{\unskip}     \fi
\ifx \showISBNxiii \undefined \def \showISBNxiii  #1{\unskip}     \fi
\ifx \showISSN     \undefined \def \showISSN      #1{\unskip}     \fi
\ifx \showLCCN     \undefined \def \showLCCN      #1{\unskip}     \fi
\ifx \shownote     \undefined \def \shownote      #1{#1}          \fi
\ifx \showarticletitle \undefined \def \showarticletitle #1{#1}   \fi
\ifx \showURL      \undefined \def \showURL       {\relax}        \fi
\providecommand\bibfield[2]{#2}
\providecommand\bibinfo[2]{#2}
\providecommand\natexlab[1]{#1}
\providecommand\showeprint[2][]{arXiv:#2}

\bibitem[\protect\citeauthoryear{??}{git}{2020}]%
        {github}
 \bibinfo{year}{2020}\natexlab{}.
\newblock \bibinfo{booktitle}{\emph{Reproduction Package}}.
\newblock
\urldef\tempurl%
\url{https://figshare.com/projects/Mining_Docker_Posts_in_SoF/81275}
\showURL{%
Retrieved May 25, 2020 from \tempurl}


\bibitem[\protect\citeauthoryear{Abdi}{Abdi}{2007}]%
        {abdi2007kendall}
\bibfield{author}{\bibinfo{person}{Herv{\'e} Abdi}.}
  \bibinfo{year}{2007}\natexlab{}.
\newblock \showarticletitle{The Kendall rank correlation coefficient}.
\newblock \bibinfo{journal}{\emph{Encyclopedia of Measurement and Statistics.
  Sage, Thousand Oaks, CA}} (\bibinfo{year}{2007}), \bibinfo{pages}{508--510}.
\newblock


\bibitem[\protect\citeauthoryear{Agrawal, Fu, and Menzies}{Agrawal
  et~al\mbox{.}}{2018}]%
        {agrawal2018wrong}
\bibfield{author}{\bibinfo{person}{Amritanshu Agrawal}, \bibinfo{person}{Wei
  Fu}, {and} \bibinfo{person}{Tim Menzies}.} \bibinfo{year}{2018}\natexlab{}.
\newblock \showarticletitle{What is wrong with topic modeling? and how to fix
  it using search-based software engineering}.
\newblock \bibinfo{journal}{\emph{Information and Software Technology}}
  \bibinfo{volume}{98} (\bibinfo{year}{2018}), \bibinfo{pages}{74--88}.
\newblock


\bibitem[\protect\citeauthoryear{Ahmed and Bagherzadeh}{Ahmed and
  Bagherzadeh}{2018}]%
        {concurrency}
\bibfield{author}{\bibinfo{person}{Syed Ahmed} {and} \bibinfo{person}{Mehdi
  Bagherzadeh}.} \bibinfo{year}{2018}\natexlab{}.
\newblock \showarticletitle{What do concurrency developers ask about? a
  large-scale study using stack overflow}. In
  \bibinfo{booktitle}{\emph{Proceedings of the 12th ACM/IEEE International
  Symposium on Empirical Software Engineering and Measurement (ESEM)}}.
  \bibinfo{pages}{1--10}.
\newblock


\bibitem[\protect\citeauthoryear{Alshangiti, Sapkota, Murukannaiah, Liu, and
  Yu}{Alshangiti et~al\mbox{.}}{2019}]%
        {whyML}
\bibfield{author}{\bibinfo{person}{Moayad Alshangiti}, \bibinfo{person}{Hitesh
  Sapkota}, \bibinfo{person}{Pradeep~K Murukannaiah}, \bibinfo{person}{Xumin
  Liu}, {and} \bibinfo{person}{Qi Yu}.} \bibinfo{year}{2019}\natexlab{}.
\newblock \showarticletitle{Why is Developing Machine Learning Applications
  Challenging? A Study on Stack Overflow Posts}. In
  \bibinfo{booktitle}{\emph{2019 ACM/IEEE International Symposium on Empirical
  Software Engineering and Measurement (ESEM)}}. IEEE, \bibinfo{pages}{1--11}.
\newblock


\bibitem[\protect\citeauthoryear{{Anderson}}{{Anderson}}{2015}]%
        {anderson2015docker}
\bibfield{author}{\bibinfo{person}{C. {Anderson}}.}
  \bibinfo{year}{2015}\natexlab{}.
\newblock \showarticletitle{Docker [Software engineering]}.
\newblock \bibinfo{journal}{\emph{IEEE Software}} \bibinfo{volume}{32},
  \bibinfo{number}{3} (\bibinfo{year}{2015}), \bibinfo{pages}{102--c3}.
\newblock


\bibitem[\protect\citeauthoryear{Asaduzzaman, Mashiyat, Roy, and
  Schneider}{Asaduzzaman et~al\mbox{.}}{2013}]%
        {asaduzzaman2013answering}
\bibfield{author}{\bibinfo{person}{Muhammad Asaduzzaman},
  \bibinfo{person}{Ahmed~Shah Mashiyat}, \bibinfo{person}{Chanchal~K Roy},
  {and} \bibinfo{person}{Kevin~A Schneider}.} \bibinfo{year}{2013}\natexlab{}.
\newblock \showarticletitle{Answering questions about unanswered questions of
  stack overflow}. In \bibinfo{booktitle}{\emph{2013 10th Working Conference on
  Mining Software Repositories (MSR)}}. IEEE, \bibinfo{pages}{97--100}.
\newblock


\bibitem[\protect\citeauthoryear{Bagherzadeh and Khatchadourian}{Bagherzadeh
  and Khatchadourian}{2019}]%
        {bigdata}
\bibfield{author}{\bibinfo{person}{Mehdi Bagherzadeh} {and}
  \bibinfo{person}{Raffi Khatchadourian}.} \bibinfo{year}{2019}\natexlab{}.
\newblock \showarticletitle{Going big: a large-scale study on what big data
  developers ask}. In \bibinfo{booktitle}{\emph{Proceedings of the 2019 27th
  ACM Joint Meeting on European Software Engineering Conference and Symposium
  on the Foundations of Software Engineering}}. \bibinfo{pages}{432--442}.
\newblock


\bibitem[\protect\citeauthoryear{Bajaj, Pattabiraman, and Mesbah}{Bajaj
  et~al\mbox{.}}{2014}]%
        {bajaj2014mining}
\bibfield{author}{\bibinfo{person}{Kartik Bajaj}, \bibinfo{person}{Karthik
  Pattabiraman}, {and} \bibinfo{person}{Ali Mesbah}.}
  \bibinfo{year}{2014}\natexlab{}.
\newblock \showarticletitle{Mining questions asked by web developers}. In
  \bibinfo{booktitle}{\emph{Proceedings of the 11th Working Conference on
  Mining Software Repositories}}. \bibinfo{pages}{112--121}.
\newblock


\bibitem[\protect\citeauthoryear{Baltes, Dumani, Treude, and Diehl}{Baltes
  et~al\mbox{.}}{2018}]%
        {sotorrent}
\bibfield{author}{\bibinfo{person}{Sebastian Baltes}, \bibinfo{person}{Lorik
  Dumani}, \bibinfo{person}{Christoph Treude}, {and} \bibinfo{person}{Stephan
  Diehl}.} \bibinfo{year}{2018}\natexlab{}.
\newblock \showarticletitle{SOTorrent: reconstructing and analyzing the
  evolution of stack overflow posts}. In \bibinfo{booktitle}{\emph{Proceedings
  of the 15th International Conference on Mining Software Repositories, {MSR}
  2018, Gothenburg, Sweden, May 28-29, 2018}}. \bibinfo{publisher}{{ACM}},
  \bibinfo{pages}{319--330}.
\newblock
\urldef\tempurl%
\url{https://doi.org/10.1145/3196398.3196430}
\showDOI{\tempurl}


\bibitem[\protect\citeauthoryear{Bangash, Sahar, Chowdhury, Wong, Hindle, and
  Ali}{Bangash et~al\mbox{.}}{2019}]%
        {whatML}
\bibfield{author}{\bibinfo{person}{Abdul~Ali Bangash}, \bibinfo{person}{Hareem
  Sahar}, \bibinfo{person}{Shaiful Chowdhury},
  \bibinfo{person}{Alexander~William Wong}, \bibinfo{person}{Abram Hindle},
  {and} \bibinfo{person}{Karim Ali}.} \bibinfo{year}{2019}\natexlab{}.
\newblock \showarticletitle{What do developers know about machine learning: a
  study of ML discussions on StackOverflow}. In \bibinfo{booktitle}{\emph{2019
  IEEE/ACM 16th International Conference on Mining Software Repositories
  (MSR)}}. IEEE, \bibinfo{pages}{260--264}.
\newblock


\bibitem[\protect\citeauthoryear{Barua, Thomas, and Hassan}{Barua
  et~al\mbox{.}}{2014}]%
        {barua2014developers}
\bibfield{author}{\bibinfo{person}{Anton Barua}, \bibinfo{person}{Stephen~W
  Thomas}, {and} \bibinfo{person}{Ahmed~E Hassan}.}
  \bibinfo{year}{2014}\natexlab{}.
\newblock \showarticletitle{What are developers talking about? an analysis of
  topics and trends in stack overflow}.
\newblock \bibinfo{journal}{\emph{Empirical Software Engineering}}
  \bibinfo{volume}{19}, \bibinfo{number}{3} (\bibinfo{year}{2014}),
  \bibinfo{pages}{619--654}.
\newblock


\bibitem[\protect\citeauthoryear{Blei, Ng, and Jordan}{Blei
  et~al\mbox{.}}{2003}]%
        {LDA}
\bibfield{author}{\bibinfo{person}{David~M Blei}, \bibinfo{person}{Andrew~Y
  Ng}, {and} \bibinfo{person}{Michael~I Jordan}.}
  \bibinfo{year}{2003}\natexlab{}.
\newblock \showarticletitle{Latent dirichlet allocation}.
\newblock \bibinfo{journal}{\emph{Journal of machine Learning research}}
  \bibinfo{volume}{3}, \bibinfo{number}{Jan} (\bibinfo{year}{2003}),
  \bibinfo{pages}{993--1022}.
\newblock


\bibitem[\protect\citeauthoryear{{Cito}, {Schermann}, {Wittern}, {Leitner},
  {Zumberi}, and {Gall}}{{Cito} et~al\mbox{.}}{2017}]%
        {cito2017dockerfiles}
\bibfield{author}{\bibinfo{person}{J. {Cito}}, \bibinfo{person}{G.
  {Schermann}}, \bibinfo{person}{J.~E. {Wittern}}, \bibinfo{person}{P.
  {Leitner}}, \bibinfo{person}{S. {Zumberi}}, {and} \bibinfo{person}{H.~C.
  {Gall}}.} \bibinfo{year}{2017}\natexlab{}.
\newblock \showarticletitle{An Empirical Analysis of the Docker Container
  Ecosystem on GitHub}. In \bibinfo{booktitle}{\emph{2017 IEEE/ACM 14th
  International Conference on Mining Software Repositories (MSR)}}.
  \bibinfo{pages}{323--333}.
\newblock


\bibitem[\protect\citeauthoryear{Ebert, Gallardo, Hernantes, and Serrano}{Ebert
  et~al\mbox{.}}{2016}]%
        {ebert2016devops}
\bibfield{author}{\bibinfo{person}{Christof Ebert}, \bibinfo{person}{Gorka
  Gallardo}, \bibinfo{person}{Josune Hernantes}, {and} \bibinfo{person}{Nicolas
  Serrano}.} \bibinfo{year}{2016}\natexlab{}.
\newblock \showarticletitle{DevOps}.
\newblock \bibinfo{journal}{\emph{{IEEE} Software}} \bibinfo{volume}{33},
  \bibinfo{number}{3} (\bibinfo{year}{2016}), \bibinfo{pages}{94--100}.
\newblock


\bibitem[\protect\citeauthoryear{Fincher and Tenenberg}{Fincher and
  Tenenberg}{2005}]%
        {fincher2005making}
\bibfield{author}{\bibinfo{person}{Sally Fincher} {and} \bibinfo{person}{Josh
  Tenenberg}.} \bibinfo{year}{2005}\natexlab{}.
\newblock \showarticletitle{Making sense of card sorting data}.
\newblock \bibinfo{journal}{\emph{Expert Systems}} \bibinfo{volume}{22},
  \bibinfo{number}{3} (\bibinfo{year}{2005}), \bibinfo{pages}{89--93}.
\newblock


\bibitem[\protect\citeauthoryear{Garg and Garg}{Garg and Garg}{2019}]%
        {garg2019automated}
\bibfield{author}{\bibinfo{person}{Somya Garg} {and} \bibinfo{person}{Satvik
  Garg}.} \bibinfo{year}{2019}\natexlab{}.
\newblock \showarticletitle{Automated Cloud Infrastructure, Continuous
  Integration and Continuous Delivery using Docker with Robust Container
  Security}. In \bibinfo{booktitle}{\emph{2019 IEEE Conference on Multimedia
  Information Processing and Retrieval (MIPR)}}. IEEE,
  \bibinfo{pages}{467--470}.
\newblock


\bibitem[\protect\citeauthoryear{Gensim}{Gensim}{2020}]%
        {gensim}
\bibfield{author}{\bibinfo{person}{Gensim}.} \bibinfo{year}{2020}\natexlab{}.
\newblock \bibinfo{booktitle}{\emph{gensim: Topic modelling for humans}}.
\newblock
\urldef\tempurl%
\url{https://radimrehurek.com/gensim/}
\showURL{%
Retrieved April 7, 2020 from \tempurl}


\bibitem[\protect\citeauthoryear{Google}{Google}{2020}]%
        {bigquery}
\bibfield{author}{\bibinfo{person}{Google}.} \bibinfo{year}{2020}\natexlab{}.
\newblock \bibinfo{booktitle}{\emph{Big Query}}.
\newblock
\urldef\tempurl%
\url{https://cloud.google.com/bigquery}
\showURL{%
Retrieved April 7, 2020 from \tempurl}


\bibitem[\protect\citeauthoryear{Hindman, Konwinski, Zaharia, Ghodsi, Joseph,
  Katz, Shenker, and Stoica}{Hindman et~al\mbox{.}}{2011}]%
        {hindman2011mesos}
\bibfield{author}{\bibinfo{person}{Benjamin Hindman}, \bibinfo{person}{Andy
  Konwinski}, \bibinfo{person}{Matei Zaharia}, \bibinfo{person}{Ali Ghodsi},
  \bibinfo{person}{Anthony~D Joseph}, \bibinfo{person}{Randy~H Katz},
  \bibinfo{person}{Scott Shenker}, {and} \bibinfo{person}{Ion Stoica}.}
  \bibinfo{year}{2011}\natexlab{}.
\newblock \showarticletitle{Mesos: A platform for fine-grained resource sharing
  in the data center.}. In \bibinfo{booktitle}{\emph{NSDI}},
  Vol.~\bibinfo{volume}{11}. \bibinfo{pages}{22--22}.
\newblock


\bibitem[\protect\citeauthoryear{Hussain and Mahmud}{Hussain and
  Mahmud}{2019}]%
        {Hussain2019pyMannKendall}
\bibfield{author}{\bibinfo{person}{Md. Hussain} {and} \bibinfo{person}{Ishtiak
  Mahmud}.} \bibinfo{year}{2019}\natexlab{}.
\newblock \showarticletitle{pyMannKendall: a python package for non parametric
  Mann Kendall family of trend tests.}
\newblock \bibinfo{journal}{\emph{Journal of Open Source Software}}
  \bibinfo{volume}{4}, \bibinfo{number}{39} (\bibinfo{date}{25 7}
  \bibinfo{year}{2019}), \bibinfo{pages}{1556}.
\newblock
\showISSN{2475-9066}
\urldef\tempurl%
\url{https://doi.org/10.21105/joss.01556}
\showDOI{\tempurl}


\bibitem[\protect\citeauthoryear{{Kang}, {Le}, and {Tao}}{{Kang}
  et~al\mbox{.}}{2016}]%
        {kang2016container}
\bibfield{author}{\bibinfo{person}{H. {Kang}}, \bibinfo{person}{M. {Le}}, {and}
  \bibinfo{person}{S. {Tao}}.} \bibinfo{year}{2016}\natexlab{}.
\newblock \showarticletitle{Container and Microservice Driven Design for Cloud
  Infrastructure DevOps}. In \bibinfo{booktitle}{\emph{2016 IEEE International
  Conference on Cloud Engineering (IC2E)}}. \bibinfo{pages}{202--211}.
\newblock


\bibitem[\protect\citeauthoryear{Le, Hin, Croft, and Babar}{Le
  et~al\mbox{.}}{2020}]%
        {le2020puminer}
\bibfield{author}{\bibinfo{person}{Triet~HM Le}, \bibinfo{person}{David Hin},
  \bibinfo{person}{Roland Croft}, {and} \bibinfo{person}{M~Ali Babar}.}
  \bibinfo{year}{2020}\natexlab{}.
\newblock \showarticletitle{PUMiner: Mining Security Posts from Developer
  Question and Answer Websites with PU Learning}.
\newblock \bibinfo{journal}{\emph{arXiv preprint arXiv:2003.03741}}
  (\bibinfo{year}{2020}).
\newblock


\bibitem[\protect\citeauthoryear{Mamykina, Manoim, Mittal, Hripcsak, and
  Hartmann}{Mamykina et~al\mbox{.}}{2011}]%
        {mamykina2011design}
\bibfield{author}{\bibinfo{person}{Lena Mamykina}, \bibinfo{person}{Bella
  Manoim}, \bibinfo{person}{Manas Mittal}, \bibinfo{person}{George Hripcsak},
  {and} \bibinfo{person}{Bj{\"o}rn Hartmann}.} \bibinfo{year}{2011}\natexlab{}.
\newblock \showarticletitle{Design lessons from the fastest q\&a site in the
  west}. In \bibinfo{booktitle}{\emph{Proceedings of the SIGCHI conference on
  Human factors in computing systems}}. \bibinfo{pages}{2857--2866}.
\newblock


\bibitem[\protect\citeauthoryear{Melo, Wiese, and D'Amorim}{Melo
  et~al\mbox{.}}{2019}]%
        {melo2019using}
\bibfield{author}{\bibinfo{person}{Luis Melo}, \bibinfo{person}{Igor~Scaliante
  Wiese}, {and} \bibinfo{person}{Marcelo D'Amorim}.}
  \bibinfo{year}{2019}\natexlab{}.
\newblock \showarticletitle{Using docker to assist Q\&A forum users}.
\newblock \bibinfo{journal}{\emph{IEEE Transactions on Software Engineering}}
  (\bibinfo{year}{2019}).
\newblock


\bibitem[\protect\citeauthoryear{(NLTK)}{(NLTK)}{2020}]%
        {nltkstopwords}
\bibfield{author}{\bibinfo{person}{Natural Language~Toolkit (NLTK)}.}
  \bibinfo{year}{2020}\natexlab{}.
\newblock \bibinfo{booktitle}{\emph{NLTK’s list of english stopwords}}.
\newblock
\urldef\tempurl%
\url{https: //gist.github.com/sebleier/554280}
\showURL{%
Retrieved April 7, 2020 from \tempurl}


\bibitem[\protect\citeauthoryear{Pahl, Brogi, Soldani, and Jamshidi}{Pahl
  et~al\mbox{.}}{2017}]%
        {pahl2017cloud}
\bibfield{author}{\bibinfo{person}{Claus Pahl}, \bibinfo{person}{Antonio
  Brogi}, \bibinfo{person}{Jacopo Soldani}, {and} \bibinfo{person}{Pooyan
  Jamshidi}.} \bibinfo{year}{2017}\natexlab{}.
\newblock \showarticletitle{Cloud container technologies: a state-of-the-art
  review}.
\newblock \bibinfo{journal}{\emph{IEEE Transactions on Cloud Computing}}
  (\bibinfo{year}{2017}).
\newblock


\bibitem[\protect\citeauthoryear{Pan, Chen, Brasileiro, Jayaputera, and
  Sinnott}{Pan et~al\mbox{.}}{2019}]%
        {pan2019performance}
\bibfield{author}{\bibinfo{person}{Yao Pan}, \bibinfo{person}{Ian Chen},
  \bibinfo{person}{Francisco Brasileiro}, \bibinfo{person}{Glenn Jayaputera},
  {and} \bibinfo{person}{Richard Sinnott}.} \bibinfo{year}{2019}\natexlab{}.
\newblock \showarticletitle{A Performance Comparison of Cloud-Based Container
  Orchestration Tools}. In \bibinfo{booktitle}{\emph{2019 IEEE International
  Conference on Big Knowledge (ICBK)}}. IEEE, \bibinfo{pages}{191--198}.
\newblock


\bibitem[\protect\citeauthoryear{Pletea, Vasilescu, and Serebrenik}{Pletea
  et~al\mbox{.}}{2014}]%
        {pletea2014security}
\bibfield{author}{\bibinfo{person}{Daniel Pletea}, \bibinfo{person}{Bogdan
  Vasilescu}, {and} \bibinfo{person}{Alexander Serebrenik}.}
  \bibinfo{year}{2014}\natexlab{}.
\newblock \showarticletitle{Security and emotion: sentiment analysis of
  security discussions on GitHub}. In \bibinfo{booktitle}{\emph{Proceedings of
  the 11th working conference on mining software repositories}}.
  \bibinfo{pages}{348--351}.
\newblock


\bibitem[\protect\citeauthoryear{Porter}{Porter}{2001}]%
        {porter}
\bibfield{author}{\bibinfo{person}{Micheal~F Porter}.}
  \bibinfo{year}{2001}\natexlab{}.
\newblock \bibinfo{booktitle}{\emph{Snowball: A language for stemming
  algorithms}}.
\newblock
\urldef\tempurl%
\url{http://snowball.tartarus.org/texts/introduction.html}
\showURL{%
Retrieved April 7, 2020 from \tempurl}


\bibitem[\protect\citeauthoryear{Puth, Neuh{\"a}user, and Ruxton}{Puth
  et~al\mbox{.}}{2015}]%
        {puth2015effective}
\bibfield{author}{\bibinfo{person}{Marie-Therese Puth}, \bibinfo{person}{Markus
  Neuh{\"a}user}, {and} \bibinfo{person}{Graeme~D Ruxton}.}
  \bibinfo{year}{2015}\natexlab{}.
\newblock \showarticletitle{Effective use of Spearman's and Kendall's
  correlation coefficients for association between two measured traits}.
\newblock \bibinfo{journal}{\emph{Animal Behaviour}}  \bibinfo{volume}{102}
  (\bibinfo{year}{2015}), \bibinfo{pages}{77--84}.
\newblock


\bibitem[\protect\citeauthoryear{R{\"o}der, Both, and Hinneburg}{R{\"o}der
  et~al\mbox{.}}{2015}]%
        {roder2015exploring}
\bibfield{author}{\bibinfo{person}{Michael R{\"o}der}, \bibinfo{person}{Andreas
  Both}, {and} \bibinfo{person}{Alexander Hinneburg}.}
  \bibinfo{year}{2015}\natexlab{}.
\newblock \showarticletitle{Exploring the space of topic coherence measures}.
  In \bibinfo{booktitle}{\emph{Proceedings of the eighth ACM international
  conference on Web search and data mining}}. \bibinfo{pages}{399--408}.
\newblock


\bibitem[\protect\citeauthoryear{Rosen and Shihab}{Rosen and Shihab}{2016}]%
        {rosen2016mobile}
\bibfield{author}{\bibinfo{person}{Christoffer Rosen} {and}
  \bibinfo{person}{Emad Shihab}.} \bibinfo{year}{2016}\natexlab{}.
\newblock \showarticletitle{What are mobile developers asking about? a large
  scale study using stack overflow}.
\newblock \bibinfo{journal}{\emph{Empirical Software Engineering}}
  \bibinfo{volume}{21}, \bibinfo{number}{3} (\bibinfo{year}{2016}),
  \bibinfo{pages}{1192--1223}.
\newblock


\bibitem[\protect\citeauthoryear{Stevens, Kegelmeyer, Andrzejewski, and
  Buttler}{Stevens et~al\mbox{.}}{2012}]%
        {stevens2012exploring}
\bibfield{author}{\bibinfo{person}{Keith Stevens}, \bibinfo{person}{Philip
  Kegelmeyer}, \bibinfo{person}{David Andrzejewski}, {and}
  \bibinfo{person}{David Buttler}.} \bibinfo{year}{2012}\natexlab{}.
\newblock \showarticletitle{Exploring topic coherence over many models and many
  topics}. In \bibinfo{booktitle}{\emph{Proceedings of the 2012 Joint
  Conference on Empirical Methods in Natural Language Processing and
  Computational Natural Language Learning}}. Association for Computational
  Linguistics, \bibinfo{pages}{952--961}.
\newblock


\bibitem[\protect\citeauthoryear{Sultan, Ahmad, and Dimitriou}{Sultan
  et~al\mbox{.}}{2019}]%
        {sultan2019container}
\bibfield{author}{\bibinfo{person}{Sari Sultan}, \bibinfo{person}{Imtiaz
  Ahmad}, {and} \bibinfo{person}{Tassos Dimitriou}.}
  \bibinfo{year}{2019}\natexlab{}.
\newblock \showarticletitle{Container Security: Issues, Challenges, and the
  Road Ahead}.
\newblock \bibinfo{journal}{\emph{IEEE Access}}  \bibinfo{volume}{7}
  (\bibinfo{year}{2019}), \bibinfo{pages}{52976--52996}.
\newblock


\bibitem[\protect\citeauthoryear{Sysdig}{Sysdig}{2020}]%
        {sysdig}
\bibfield{author}{\bibinfo{person}{Sysdig}.} \bibinfo{year}{2020}\natexlab{}.
\newblock \bibinfo{booktitle}{\emph{2018 docker usage report.}}
\newblock
\urldef\tempurl%
\url{https://sysdig.com/blog/2018-docker-usage-report/}
\showURL{%
Retrieved April 7, 2020 from \tempurl}


\bibitem[\protect\citeauthoryear{Tan, Wang, and Lee}{Tan et~al\mbox{.}}{2002}]%
        {tan2002use}
\bibfield{author}{\bibinfo{person}{Chade-Meng Tan}, \bibinfo{person}{Yuan-Fang
  Wang}, {and} \bibinfo{person}{Chan-Do Lee}.} \bibinfo{year}{2002}\natexlab{}.
\newblock \showarticletitle{The use of bigrams to enhance text categorization}.
\newblock \bibinfo{journal}{\emph{Information processing \& management}}
  \bibinfo{volume}{38}, \bibinfo{number}{4} (\bibinfo{year}{2002}),
  \bibinfo{pages}{529--546}.
\newblock


\bibitem[\protect\citeauthoryear{Tian, Ng, Cao, and McIntosh}{Tian
  et~al\mbox{.}}{2019}]%
        {cmc.2019.07818}
\bibfield{author}{\bibinfo{person}{Yijun Tian}, \bibinfo{person}{Waii Ng},
  \bibinfo{person}{Jialiang Cao}, {and} \bibinfo{person}{Suzanne McIntosh}.}
  \bibinfo{year}{2019}\natexlab{}.
\newblock \showarticletitle{Geek Talents: Who are the Top Experts on GitHub and
  Stack Overflow?}
\newblock \bibinfo{journal}{\emph{Computers, Materials \& Continua}}
  \bibinfo{volume}{61}, \bibinfo{number}{2} (\bibinfo{year}{2019}),
  \bibinfo{pages}{465--479}.
\newblock
\showISSN{1546-2226}
\urldef\tempurl%
\url{https://doi.org/10.32604/cmc.2019.07818}
\showDOI{\tempurl}


\bibitem[\protect\citeauthoryear{Tutorial}{Tutorial}{2020}]%
        {mysql}
\bibfield{author}{\bibinfo{person}{Tutorial}.} \bibinfo{year}{2020}\natexlab{}.
\newblock \bibinfo{booktitle}{\emph{PostGreSQL vs MySQL.}}
\newblock
\urldef\tempurl%
\url{https://www.postgresqltutorial.com/postgresql-vs-mysql/}
\showURL{%
Retrieved April 7, 2020 from \tempurl}


\bibitem[\protect\citeauthoryear{{Velásquez}, {Munoz–Arcentales}, and
  {Rodriguez}}{{Velásquez} et~al\mbox{.}}{2018}]%
        {8441979}
\bibfield{author}{\bibinfo{person}{W. {Velásquez}}, \bibinfo{person}{A.
  {Munoz–Arcentales}}, {and} \bibinfo{person}{J.~S. {Rodriguez}}.}
  \bibinfo{year}{2018}\natexlab{}.
\newblock \showarticletitle{A Case Study: Ingestion Analysis of WSN Data in
  Databases using Docker}. In \bibinfo{booktitle}{\emph{2018 1st International
  Conference on Computer Applications Information Security (ICCAIS)}}.
  \bibinfo{pages}{1--6}.
\newblock


\bibitem[\protect\citeauthoryear{Xu and Marinov}{Xu and Marinov}{2018}]%
        {xu2018mining}
\bibfield{author}{\bibinfo{person}{Tianyin Xu} {and} \bibinfo{person}{Darko
  Marinov}.} \bibinfo{year}{2018}\natexlab{}.
\newblock \showarticletitle{Mining container image repositories for software
  configuration and beyond}. In \bibinfo{booktitle}{\emph{Proceedings of the
  40th International Conference on Software Engineering: New Ideas and Emerging
  Results}}. \bibinfo{pages}{49--52}.
\newblock


\bibitem[\protect\citeauthoryear{Yang, Lo, Xia, Wan, and Sun}{Yang
  et~al\mbox{.}}{2016}]%
        {yang2016security}
\bibfield{author}{\bibinfo{person}{Xin-Li Yang}, \bibinfo{person}{David Lo},
  \bibinfo{person}{Xin Xia}, \bibinfo{person}{Zhi-Yuan Wan}, {and}
  \bibinfo{person}{Jian-Ling Sun}.} \bibinfo{year}{2016}\natexlab{}.
\newblock \showarticletitle{What security questions do developers ask? a
  large-scale study of stack overflow posts}.
\newblock \bibinfo{journal}{\emph{Journal of Computer Science and Technology}}
  \bibinfo{volume}{31}, \bibinfo{number}{5} (\bibinfo{year}{2016}),
  \bibinfo{pages}{910--924}.
\newblock


\bibitem[\protect\citeauthoryear{Zahedi, Rajapakse, and Babar}{Zahedi
  et~al\mbox{.}}{2020}]%
        {zahedi2020devops}
\bibfield{author}{\bibinfo{person}{Mansooreh Zahedi},
  \bibinfo{person}{Roshan~Namal Rajapakse}, {and} \bibinfo{person}{Muhammad~Ali
  Babar}.} \bibinfo{year}{2020}\natexlab{}.
\newblock \showarticletitle{Mining Questions Asked about Continuous Software
  Engineering: A Case Study of Stack Overflow}. In
  \bibinfo{booktitle}{\emph{Proceedings of the Evaluation and Assessment in
  Software Engineering}} (Trondheim, Norway) \emph{(\bibinfo{series}{EASE
  ’20})}. \bibinfo{publisher}{Association for Computing Machinery},
  \bibinfo{address}{New York, NY, USA}, \bibinfo{pages}{41–50}.
\newblock
\showISBNx{9781450377317}
\urldef\tempurl%
\url{https://doi.org/10.1145/3383219.3383224}
\showDOI{\tempurl}


\bibitem[\protect\citeauthoryear{{Zeng}, {Wang}, {Deng}, and {Zhang}}{{Zeng}
  et~al\mbox{.}}{2017}]%
        {8250343}
\bibfield{author}{\bibinfo{person}{H. {Zeng}}, \bibinfo{person}{B. {Wang}},
  \bibinfo{person}{W. {Deng}}, {and} \bibinfo{person}{W. {Zhang}}.}
  \bibinfo{year}{2017}\natexlab{}.
\newblock \showarticletitle{Measurement and Evaluation for Docker Container
  Networking}. In \bibinfo{booktitle}{\emph{2017 International Conference on
  Cyber-Enabled Distributed Computing and Knowledge Discovery (CyberC)}}.
  \bibinfo{pages}{105--108}.
\newblock


\end{thebibliography}


\end{document}